\documentclass[lettersize,journal]{IEEEtran}
\usepackage{amsmath,amsfonts}
\usepackage{algorithmic}
\usepackage{array}
\usepackage{flafter}
\usepackage{afterpage}
\usepackage[caption=false,font=normalsize,labelfont=sf,textfont=sf]{subfig}
\usepackage{textcomp}
\usepackage{stfloats}
\usepackage{url}
\usepackage{placeins}
\usepackage{verbatim}
\usepackage{graphicx}
\usepackage{amsmath,amsfonts}
\usepackage{algorithmic}
\usepackage{tabularx}
\usepackage{array}
\usepackage{xcolor}
\usepackage{multirow}
\pdfoutput=1

\def\BibTeX{{\rm B\kern-.05em{\sc i\kern-.025em b}\kern-.08em
    T\kern-.1667em\lower.7ex\hbox{E}\kern-.125emX}}
\usepackage{balance}

\begin{document}
\title{Streaming quanta sensors for online, high-performance imaging and vision}
\author{Tianyi Zhang, Matthew Dutson, Vivek Boominathan, Mohit Gupta, Ashok Veeraraghavan
\thanks{Tianyi Zhang, Vivek Boominathan, Ashok Veeraraghavan are with the ECE department at Rice University, USA; \\
Matthew Dutson and Mohit Gupta are with the CS department of University of Wisconsin-Madison, USA; \\
\textit{Corresponding author: Ashok Veeraraghavan}
}}



\maketitle
\begin{abstract}
Recently quanta image sensors (QIS) -- ultra-fast, zero-read-noise binary image sensors-- have demonstrated remarkable imaging capabilities in many challenging scenarios. Despite their potential, the adoption of these sensors is severely hampered by (a) high data rates and (b) the need for new computational pipelines to handle the unconventional raw data. We introduce a simple, low-bandwidth computational pipeline to address these challenges. Our approach is based on a novel streaming representation with a small memory footprint, efficiently capturing intensity information at multiple temporal scales. Updating the representation requires only 16 floating-point operations/pixel, which can be efficiently computed online at the native frame rate of the binary frames. We use a neural network operating on this representation to reconstruct videos in real-time (10-30 fps). We illustrate why such representation is well-suited for these emerging sensors, and how it offers low latency and high frame rate while retaining flexibility for downstream computer vision. Our approach results in significant data bandwidth reductions ($\sim100\times$) and real-time image reconstruction and computer vision --$10^4-10^5 \times$)reduction in computation than existing state-of-the-art approach\cite{ma2020quanta}, while maintaining comparable quality. To the best of our knowledge, our approach is the first to achieve online, real-time image reconstruction on QIS.
\end{abstract}

\section{Introduction}

\IEEEPARstart{U}{ltra-fast} binary quanta image sensors (QIS) have demonstrated novel imaging capabilities in challenging environments, including high dynamic range imaging, ultra-low light imaging, and high-speed imaging \cite{ma2020quanta}\cite{fossum2016quanta}\cite{ingle2019high}. This makes them attractive for a wide range of applications such as autonomous driving, robotics, remote sensing, and bio-imaging, where low light, large brightness changes, and/or fast motion are prevalent.

QIS sensors work by coupling single photon detectors with ultra-fast frame readout circuits. SPAD-QIS, a prominent QIS implementation \cite{ulku2018512}\cite{duttonSPADbasedQVGA2016}, can output binary frames free of read noise at speeds reaching up to 100~kFPS. The binary data streams generated by these sensors contain rich, fine-grained temporal information. Recent works have shown that when given access to all the binary data, it is possible to align and fuse the binary frames to output clean and sharp image/video reconstructions, allowing us to resolve high-speed details, even under low light levels and high dynamic range \cite{ma2020quanta}\cite{chandramouliBitToo2019}.

\textbf{Challenges of quanta sensing:} While quanta sensors are gradually becoming more readily available, there are several critical challenges that need to be addressed before they can be deployed in real-world applications. 
The first challenge is the \textit{huge data bandwidth} needed to stream the raw binary frames off the sensor. 
A QIS of 1~megapixel operating at 100~kFPS produces data at a rate of 100~Gbps---an impractically high-bandwidth data stream. This large bandwidth also results in prohibitive power and memory requirements. The second problem is the large computational cost involved with processing all the binary frames to produce quality images or perform downstream inferences. This problem is exacerbated by the unconventional nature of raw sensor data (a binary, quantized, and noisy bit-stream), which necessitates additional pre-processing steps before traditional computer vision algorithms can be applied. Recent pre-processing algorithms \cite{ma2020quanta} require several minutes of processing per second of input data---a computational cost that is four to five orders of magnitude slower than real-time. In spite of the promise of quanta image sensors, these practical challenges are inhibiting the deployment of this promising technology in resource-constrained applications that need to operate in real time.

\begin{figure*}[!htbp]
    \centering
    \includegraphics[width=1.0\linewidth]{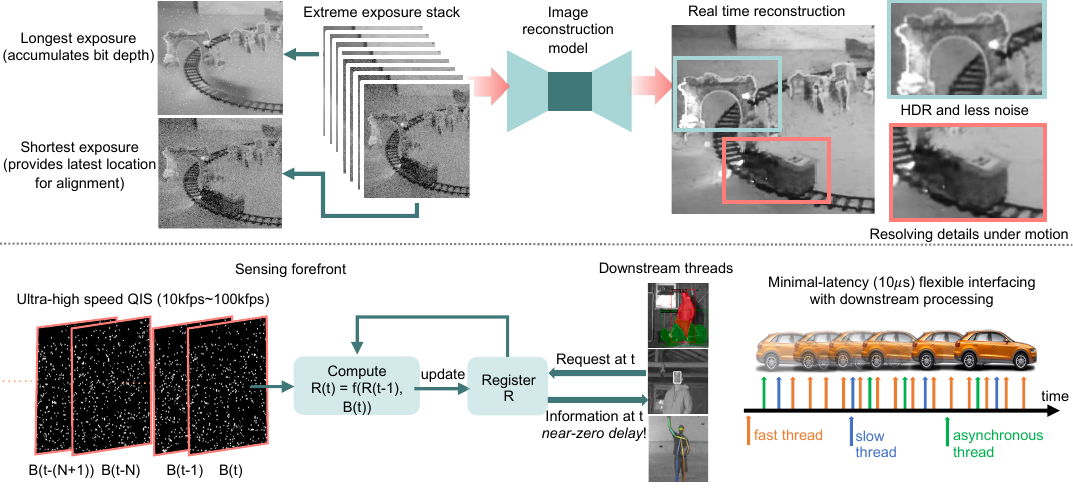}
    \caption{\textbf{Top:} We propose a novel online processing architecture for QIS, which consists of streaming an extreme multi-exposure stack, and a modified U-Net with ResNet blocks for performing the reconstruction. The approach results in good image reconstruction performance. \textbf{Bottom:} The exposure stack is computed via a streaming update (figure bottom). Such a representation ensures information up to the most recent time instant is always available and decouples exposure from frame-read out. The sensor produces a consistent and continuously updated multi-exposure stack for any downstream request with near-zero lag.}
    \label{fig:teaser}
\end{figure*}

We propose data structures, computational architectures, and associated algorithms that together address these challenges and close the gap towards near-real-time operation with QIS sensors. First, we propose a compact streaming representation that has a low memory footprint, can be updated in real-time with low compute complexity (a small number of additions and multiplications per pixel update), and stores rich information about the scene texture and motion at multiple temporal resolutions. We couple this representation with a feed-forward neural network to yield image reconstruction and high-level inference in real time (Fig.~\ref{fig:teaser}), with quality comparable to existing, resource-intensive approaches \cite{ma2020quanta}. Together, this representation and a feed-forward reconstruction algorithm yield near-real-time performance (Fig.~\ref{fig:teaser}), with a significant reduction in data bandwidth.

\textbf{Streaming representations:} The proposed streaming representations are updated when every new QIS frame arrives, thus minimizing the latency between the occurrence of a change and its capture. These high-speed updates are only possible due to the lack of read noise in SPAD-QIS; in a conventional sensor, sampling the scene at an equivalent rate would result in unacceptably high read noise.

Our streaming representation maintains information at multiple temporal scales, thereby encoding information about objects moving at varying speeds. Downstream inference algorithms (e.g., image reconstruction, face tracking, object detection) can \textit{poll} the representation at an arbitrary rate, based on the application needs and the computational resources available. Our representation allows us to build a seamless interface between a $\approx$100~kFPS sensor and $\approx$100~FPS vision algorithms, without incurring a latency penalty.

\textbf{Contributions:}
In summary, our proposed pipeline enables high-speed streaming vision with QIS sensors by combining a compact, streaming representation with a feed-forward neural network. The main contributions of the paper are:
\begin{itemize}
    \item A novel online representation and corresponding processing algorithms for quanta image sensors that are suitable for real-time systems.
    This representation lowers the bandwidth requirements by approximately two orders of magnitude (relative to reading out raw bitplanes), to a level compatible with current processing pipelines and data transfer interfaces.
    This representation can be continuously updated with just 16 floating point operations per pixel per binary frame, resulting in low latency and high computational efficiency.
    \item We design a feed-forward neural network for image reconstruction that denoises, aligns and integrates the multi-exposure representation into a clean intensity frame. The network is compact and can run in real-time on a desktop GPU (tens of FPS), resulting in an 4-5 orders of magnitude speed-up compared to existing methods for image reconstruction from quanta sensors. 
    \item We provide a rich, semi-realistic dataset that combines data from real cameras with 3D-simulated data. This dataset can be used for supervised learning on dynamic QIS problems. It contains diverse object categories and motion statistics. It generalizes across light levels, and can realistically represent camera captures of the 3D world.
    \item We demonstrate the application of our representation, architecture, and algorithms to several traditional computer vision tasks such as face detection, object detection, human pose estimation, and tracking. We demonstrate for the first time that such computer vision tasks can be applied to QIS data with near-real-time performance. 
\end{itemize}

\section{Literature Review}
\subsection{Image processing and vision with QIS}
Single-photon QIS sensors output binary frames with a high temporal resolution, up to 100~kFPS, with zero read noise due to the avalanche process \cite{ulku2018512,duttonSPADbasedQVGA2016}. Early works in this field study processing methods suitable for static scenes and demonstrated that QIS is especially effective for resolving low light and high dynamic range (HDR) scenes \cite{yangBitsPhotons2012, chanImagesBits2016, choiImageReconstruction2018,gnanasambandam2019megapixel, ingle2019high}.

The ability of QIS to access information at high temporal resolution can allow us to overcome the noise-blur tradeoff that exists in conventional single-exposure sensors. However, it is challenging to accurately align and merge binary frames under large motion, due to the high shot noise in individual binary frames. A few works compensate for simple motion by shifting and fusing the binary frames to produce a single, blur-free image reconstruction \cite{fossum2013modeling,gyongy2017object,iwabuchi2019iterative}. However, such methods work only for simple cases such as rigid objects with planar motion, and have difficulty handling complex cases such as occlusion. Seets et al. \cite{seets2021motion} use flux and motion change points to guide the integration process, in order to minimize motion blur. Recently, Quanta Burst Photography (QBP) \cite{ma2020quanta} has demonstrated more robust performance across realistic and complex scenes. QBP assumes that the motion within a sequence can be approximated using patch-wise optical flow. However, the method requires full access to all the raw binary data and can easily take minutes or more to compute, rendering it difficult to run in real-time. Furthermore, the QBP alignment step requires explicit optical flow guidance, which can be unstable and fail at longer time horizons. Chandramouli et al. \cite{chandramouli2019bit} proposed a learning-based approach for denoising binary sequences, but still assume reading out all raw data.
\vspace{-1em}
\subsection{QIS and SPAD datasets for supervised learning}
Capturing paired binary sequence datasets, i.e., binary frames with the ground truth intensity, is challenging in high-speed and low-light conditions. In low-light, static environments, it is possible to obtain paired single-frame data by capturing sets of short and long exposure shots, similar to the method used to create the SID dataset \cite{chen2018learning}. However, this approach is infeasible when operating at the speed and duration of QIS. For commercial high-speed sensors with full bit-depth capacities, read noise can reach up to 10e- rms, resulting in extremely low SNR under moderate or low light.

Due to such difficulties in real captures, some works attempt to synthesize paired datasets. \cite{bian2022large} produced a large single-photon dataset for single-image super-resolution by accurately approximating the SPAD sensor characteristics. \cite{chi2020dynamic} synthesized dynamic sequences by considering global 2D image translations and translating patches on top of the static background to mimic local motion. However, this simulated dataset does not faithfully capture effects such as self-occlusion and spatially varying motion caused by depth differences. Chandramouli et al. \cite{chandramouli2019bit} interpolated 240~FPS video sequences to mimic high-speed captures and applied quantization. This work inspired our data generation approach. 

\vspace{-1em}
\subsection{Vision with QIS under sparse photon arrivals}
It has been shown that QIS can enable low-light vision with less than 1 photon per pixel, for tasks such as classification~\cite{gnanasambandam2020image,goyal2021photon}, object detection~\cite{li2021photon}, and depth prediction~\cite{goyal2021photon}. These works show that visual information can be uncovered with very few photons. For dynamic scenes, Ma et al. \cite{ma2023burst} show the promise of QIS for computer vision, by applying off-the-shelf networks for various tasks by first reconstructing images from binary data.
\vspace{-1em}
\subsection{Streaming algorithms and perception}
Streaming algorithms \cite{gilbert2001surfing} are designed to handle fast data streams in a real-time and one-pass manner. As an example, streaming perception \cite{li2020towardsstreaming} handles streams of fast-changing visual information so that the perception pipeline keeps up with the most recent state in the world. This concept inspired our streaming quanta sensor approach, with the goal of keeping up with the latest intensity information of a dynamic scene for minimum sensing latency.
\vspace{-1em}
\subsection{Multi-exposure architectures and algorithms}
Several high dynamic range (HDR) vision sensors capture a stack of different exposures sequentially, to handle objects and regions with drastically different light levels and speeds. Another relevant sensor architecture is the multi-bucket sensor \cite{wan2012cmos}, where multiple buckets are attached to each pixel, each capturing a different exposure at the same time. The multi-bucket sensor reduces ghosting and shortens the time to capture an exposure stack. 

While many multi-exposure fusion algorithms have been developed \cite{xu2022multi}, these works are typically specific to their own capture configurations and are not directly applicable to QIS exposures. Further, these methods do not consider a streaming setting, and assume a high enough photon level that high-flux regions in the short exposures can be directly fused to the final photograph after denoising. This assumption does not hold for QIS, where short exposures contain heavy shot noise.

\section{Quanta Streaming Representation}
We design our QIS representation such that it satisfies the following properties:

\begin{figure*}[!hbt]
    \centering
    \includegraphics[width=1.0\linewidth]{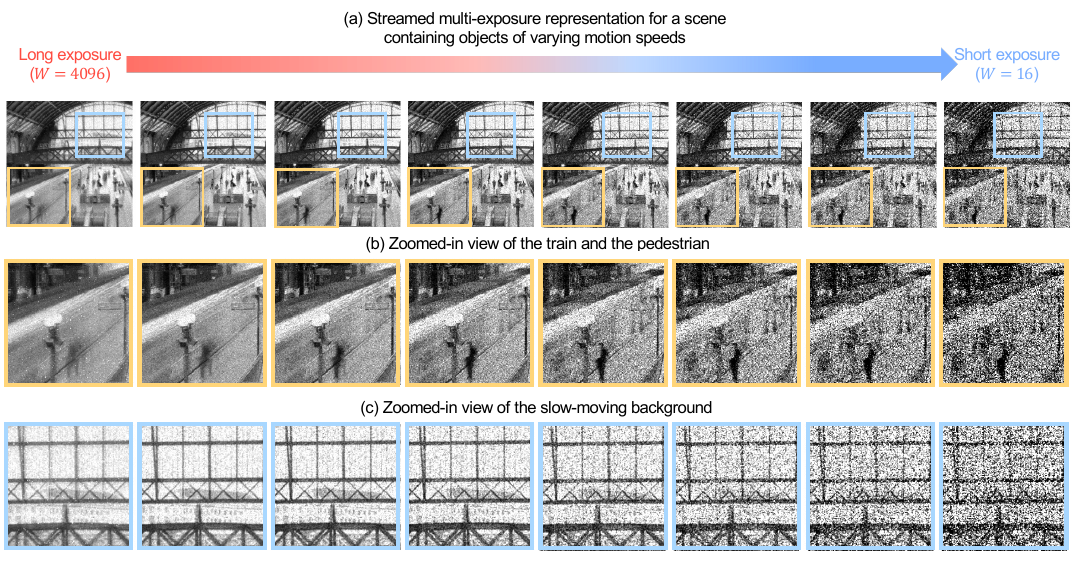}
    \caption{The streamed multi-exposure set. \textbf{(a)} shows our streamed multi-exposure representation (top left: long exposure, bottom right: short exposure; read row by row), in a scene containing objects with very different motion speeds. \textbf{(b)} shows a zoomed-in view of the train and a pedestrian. The train moves at a blazingly fast speed, while the pedestrian moves at a slower but still significant speed. The shortest exposure provides contour information for localization and alignment (e.g., windows/doors of the train, contour of the pedestrian) but lacks detail due to significant noise. The longer exposures reveal more information about the motion trajectory/speed and provide more bit-depth for resolving intensities. (c) shows a zoomed-in view of the window, which has high flux and lower contrast. There is very mild motion due to camera motion. Long/medium exposures can be directly used to resolve details in such cases (e.g., the building in the far background).}
    \label{fig:exposure_set}
\end{figure*}

\begin{itemize}
\item{\textbf{Multi-exposure representation:} Quanta sensors operate at about 10--100~kFPS natively, but any subsequent representation (or ``sketch'') should summarize the spatiotemporal volume at multiple timescales simultaneously. This will allow the representation to provide the optimal tradeoff between noise and motion blur, regardless of object velocity. To support a wide range of applications, the virtual exposure set needs to traverse a large range of exposures, covering, for example, a virtual exposure of both $\approx$100 microseconds and $\approx$10 milliseconds.} 
\item{\textbf{Ultra-fast update:} The representation should be updated per every quanta frame, i.e., approximately every 10 microseconds. This high update rate implies that the representation needs to be updated with only a few arithmetic operations per pixel per quanta frame.}
\item{\textbf{Low memory footprint:} The representation needs to be compact and should allow us to achieve a $\approx$100x reduction in memory compared to the raw quanta data.}
\end{itemize}
\vspace{-1em}
\subsection{Streaming multi-exposure mean representation}
We propose an \textit{online multi-exposure mean} representation---by generating an exponentially integrated set of exposures. 
The mean with a faster decay (negative exponential coefficient with a larger magnitude) provides us with noisy image information but captures high-speed dynamics with low motion blur. In contrast, the mean with a slower decay provides us with high SNR image information, albeit potentially with motion blur.
The rationale for choosing an exponentially-integrated mean (as opposed to, e.g., a windowed mean), is that efficient ``online'' or streaming algorithms exist \cite{gilbert2001surfing} that allow updating the exponential mean with only one multiplication and one addition per pixel per update step.
Finally, storing each of these exponentially-weighted means as an 8-bit image allows us to achieve a low memory footprint, equivalent to storing a small number of conventional images.

\textbf{How QIS senses bits and photons:}
SPAD-QIS operates at 10--100~kFPS. Within each cycle, the QIS generates a binary measurement, indicating whether a photon is present in the cycle. SPADs cannot detect more than 1 photon per cycle, and hence have a ``soft-saturating'', non-linear forward model \cite{ma2020quanta}\cite{ingle2019high}. Let $t \in N+$ be the cycle number. Let $B(t)$ be the binary measurement for the cycle. The probability of one or more photons arriving at each cycle is $P(t) = 1 - \exp(-\lambda(t))$. $B(t)$ follows $\text{Bernoulli}(p)$. Given a stack of $N$ binary frames, the sum of counts follow $\text{Binomial}(N, p)$.

\textbf{Exponential mean representation or ``sketch'':} 
The exponential (aged mean) representation is an approximation to the standard mean representation, that allows streaming updates with little compute and memory. Instead of applying a box filter on a time series, an infinite, exponential-shaped filter is applied. To compute the aged mean of the binary stream, we use the following recursion:
\begin{equation}
   R(t) = (1 - \alpha) R(t-1) + \alpha B(t)
\end{equation}

Here $R(t)$ represents the exponential mean at time t. $\alpha$ is termed the aging factor. The greater the value of $\alpha$, the more heavily recent samples are weighted. Let $W = 1 / \alpha$. Effectively, $W$ can be considered as the window or exposure parameter---the greater the value of $\alpha$, the shorter the effective exposure window. We construct a stack of time-decaying exposures using different $\alpha$ values, e.g., an exponential set as typically used in exposure bracketing. In this work, we choose $W = 2^{12}, 2^{10}, ..., 2^4$, with a maximum window $W_\text{max} =$~4096 and a minimum window $W_\text{min} =$~16. These window values allow us to handle a range of fluxes and motion speeds.

\textbf{Practical considerations:} A single exponent approximately covers a certain time of integration. Multiple exponents allow multiple temporal resolutions, but $N$ exponents require $N$ times more memory. Each frame is binary, so the smallest effective integration period should be no smaller than about 16 binary frames (1.6 milliseconds at 10~kFPS, 160 microseconds at 100~kFPS). The largest exposure is 256 longer than the shortest exposure. Together they cover the range of object velocities from ultra-fast to normal human motion while keeping the memory footprint only $8\times$ that of a conventional sensor (and $\approx 100\times$ smaller than raw quanta data).
\vspace{-1em}
\subsection{Advantages of the compact streaming representation}
Streaming representations and computational processing pipelines that operate in real-time on the most up-to-date data, have been shown to be critical for embodied perception applications such as robotics, autonomous navigation, and machine vision \cite{li2020towardsstreaming}.
The key rationale is that in embodied systems, there is always a lag between the current state of the world and the state as estimated by the perception stack. This lag is due both to representations being out-of-date and to subsequent processing times being finite.
This lag causes significant errors in actions taken by embodied agents, as these actions are based on an outdated estimate of the state of the world.
Streaming representations, processing algorithms, and error metrics (which explicitly account for this lag) result in more stable and better-performing embodied systems.
Inspired by this argument \cite{li2020towardsstreaming},  we develop an ultra-fast streaming representation that is updated with near-zero-lag (less than 10 microseconds), and a real-time processing chain for image/video reconstruction and other downstream computer vision tasks.

Streaming representations and processing pipelines on quanta sensors provide significant benefits over competing processing architectures as detailed below. 

\textbf{Sensing and processing chains decoupled:}
A streaming representation naturally allows the sensing and the subsequent post-processing pipelines to be decoupled, allowing both to be independently optimized.
In particular, downstream image/video reconstruction and other computer vision algorithms can be designed and run with application-dependent frame rates independent of the sensor sampling rate, while still ensuring that the downstream algorithms have access to the most up-to-date data.
Imagine the same quanta sensor, being used for two different applications: face detection ($\approx$10~FPS) and robotic vision ($\approx$1000~FPS). 
The streaming representations allow both of these algorithms to simultaneously access the most up-to-date data from the sensor at completely independent and asynchronous rates, without additional overhead.

\begin{figure*}[!hbpt]
    \centering
    \includegraphics[width=\linewidth]{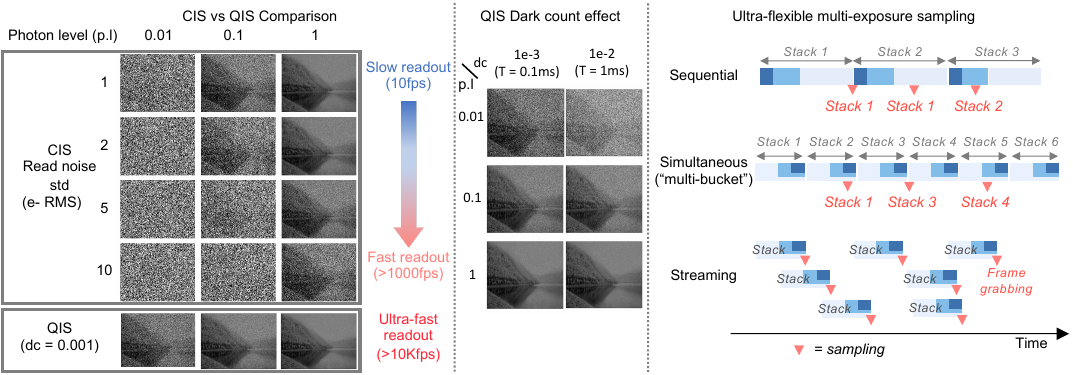}
    \caption{Why are streaming representations suitable for quanta sensors? (dc = dark counts per exposure, p.l = photon level, in number of photons). \textbf{Left column}: Conventional cameras suffer from read noise that increases with higher readout rates. This obscures scene details under low photon levels. QIS technology, on the other hand, has no read noise at any readout rate, making it ideal for preserving details during short exposures and in streaming sensors. Here the assumption is that $\text{QE} = 1$ for both sensors, and dark count rate to be $\approx$ 10cps. \textbf{Middle column}: 
    Current QIS devices have relatively low dark counts, such that even in a longer exposure (more dark counts per exposure) with sparse photon counts, the features can still be visible in the captured frames. \textbf{Right column}: We examine frame delay and latency in different multi-exposure architectures. Here the darker bars represent shorter exposures. Conventional approaches cause lag and inefficiencies for downstream processing and require precise synchronization. (Row 1) Sequential exposure capturing leads to long capture times per exposure stack. (Row 2) The multi-bucket model captures multiple exposures at once, but the discretization issue persists. (Row 3) Computing multi-exposures in a streaming manner provides the latest information at any time, reducing downstream inefficiencies.}
    \label{fig:whystream}
\end{figure*}

\textbf{Frame-based sensing latency:} 
A frame-based representation as is common (and unavoidable in conventional sensors) results in sensing latency that is on average about half the frame duration. On conventional 30--60~FPS sensors, this translates to a sensing latency of about 15--30~ms. This sensing latency increases linearly with total capture time---a particular issue in scenes with high noise or dynamic range that require longer exposures or multiple frames. In such scenarios, the latency could approach 1 second. With a multi-time-resolution streaming representation that continuously updates every 10~$\mu$s as proposed in this paper, the sensing latency is significantly reduced. This idea is illustrated in the right column of Fig.~\ref{fig:whystream}. Conventional framed-based methods require integration over a fixed duration. Thus, downstream components can only access information sampled at discrete time intervals. The streaming implementation allows seamless sampling of the exposure stack at any time, with latency as low as 10~$\mu$s.

\textbf{Read-noise advantages:}
In practice, a streaming representation also has a finite, frame-like update speed; however, this update speed is much much faster ($\approx$10~$\mu$s) than conventional sensors ($\approx$10~ms).
Does this mean that such a streaming representation is equally suited for conventional CIS (non-QIS) sensors? Unfortunately, conventional sensors have significant per-measurement read noise; as a consequence, arbitrarily reducing the frame rate results in significant noise amplification. This effect is shown in the left column of Fig.~\ref{fig:whystream}. As the read rate or update rate increases, the read noise also significantly increases, such that it overwhelms fine intensity variations at a fine time scale. Fortunately, quanta sensors inherently have no read-noise. This makes streaming representations a natural option. Fine intensity information can be captured with only shot-noise present. While SPADs have dark count effects (middle column of Fig.~\ref{fig:whystream}), fine visual information is visible even if photon arrivals are sparse.

\textbf{Streaming accuracy:} 
Finally, as argued in \cite{li2020towardsstreaming}, in practice, the accuracy that matters for embodied agents that interact with the real world is ``streaming accuracy.'' A streaming representation along with real-time post-processing algorithms allows us to optimize streaming accuracy for such systems.

\vspace{-1em}
\subsection{Visual Analysis of the Streaming Representation}
The streaming multi-exposure representation stores information about the photon spatiotemporal cubes at multiple time scales, allowing us to resolve objects with a large range of velocities, as shown in Fig.~\ref{fig:exposure_set}. Long exposures store sufficient bit-depth for resolving high dynamic range and noise, but are only suitable for slow-moving objects. Short exposures, although noisy, capture important semantic and location information about fast-moving objects. The exposure set provides photon integration and motion information at multiple time scales, allowing us to resolve a large range of motion speeds and flux intensities. In Fig.~\ref{fig:exposure_set}~(c),  low-speed objects (background) can be directly captured with a high to medium exposure. For fast-moving objects  ((b)---a very fast train and a medium-fast pedestrian), short- and medium-duration exposures can be used to resolve motion.

\section{Learning-based image/video reconstruction}
We use a U-Net architecture for image reconstruction, which allows high-quality reconstruction from the multi-exposure set. In the first part of this section, we discuss the architecture, loss design, etc. In the second part, we discuss how to systematically create synthetic datasets for training.
\vspace{-1em}
\subsection{Image reconstruction neural network}
We use a neural network to reconstruct the latest photon flux from the extreme multi-exposure set. The feed-forward network can achieve real-time performance and give high-quality reconstructions due to its effectiveness in learning from a large dataset. Our reconstruction network is based on the U-Net architecture~\cite{ronneberger2015u}, which applies multi-resolution feature extraction with skip connections between encoder and decoder blocks of the same scale. We modify the basic U-Net architecture; at each scale, we use a residual block~\cite{he2016deep, zhang2018road} with two convolution layers of kernel size $3\times3$. A pixel-shuffle convolution layer~\cite{shi2016real} is used in the encoder for downsampling, while an interpolation-based upsampling is used in the decoder. The inputs correspond to the 8-channel extreme multi-exposure set, while the output is a single-channel grayscale image. We train the network to reconstruct an image that is temporally aligned with the latest time instant or the state represented in the shortest exposure. 
For low light, we minimize a combined loss with L1 and L1 on image gradients in x and y directions. Specifically:
\begin{align}
    \text{Loss}(\lambda, \lambda_{gt}) = &\text{L1}(\lambda, \lambda_{gt}) +\sigma \text{L1}(\nabla_{x} \lambda, \nabla_{x} \lambda_{gt}) \\
    &+ \sigma \text{L1}(\nabla_{y} \lambda, \nabla_{y} \lambda_{gt})
\end{align}

Where $\lambda_{gt}$ stands for the ground truth frame (photon fluxes), that temporally aligns with the latest binary frame captured in the stack. The gradient terms are important for preserving the high-frequency components and fine details in the images. We choose a value of 0.1 for $\sigma$ -- which corresponds to the weightage value for the gradient terms. 

In real scenes, the photon flux can range dramatically from tens of photons per second to more than 10k photons per second. To enable the network to learn from a wide range of fluxes, we minimize the loss on a tone-mapped version of the intensity $\lambda$, which we denote as $T(\lambda)$. For $T$, we employ a differentiable $\mu$-law function as follows:
\begin{equation}
    T(\lambda) =  \frac{\log(\text{ReLU}(1+\mu \lambda) + \zeta)}{\log(1+\mu)}
\end{equation}
We choose a high compression rate $\mu =10^3$. In some cases, the network may predict a negative value for $\lambda$. We use the ReLU operation and the stability parameter $\zeta = 10^{-7}$ to prevent negative or zero arguments to the log function. 
\vspace{-1em}
\subsection{QIS dynamic dataset}
In this section, we discuss how to obtain large-scale datasets of QIS data so that image/video reconstruction algorithms and downstream computer vision applications can be appropriately trained. For supervised training, we aim to create diverse and realistic sequences suitable for high-speed SPAD-QIS. In particular, the dataset needs to satisfy a few requirements:
(1) the sequences must contain fine-grained inter-frame motion that can be used to simulate a slow-motion capture on the order of 10--100~kFPS
(2) the dataset should be of considerable size, high quality and largely free of artifacts
(3) the dataset should cover a range of motion levels and several types of motion, cover a large range of light levels and dynamic ranges that are close to the real world, and contain diverse textures and object types
(4) the dataset should contain realistic simulations of the sensor characteristics such as dark counts and hot pixels.

In this section, we describe the creation of such a dataset.

\begin{figure}
    \centering
    \includegraphics[width=\linewidth]{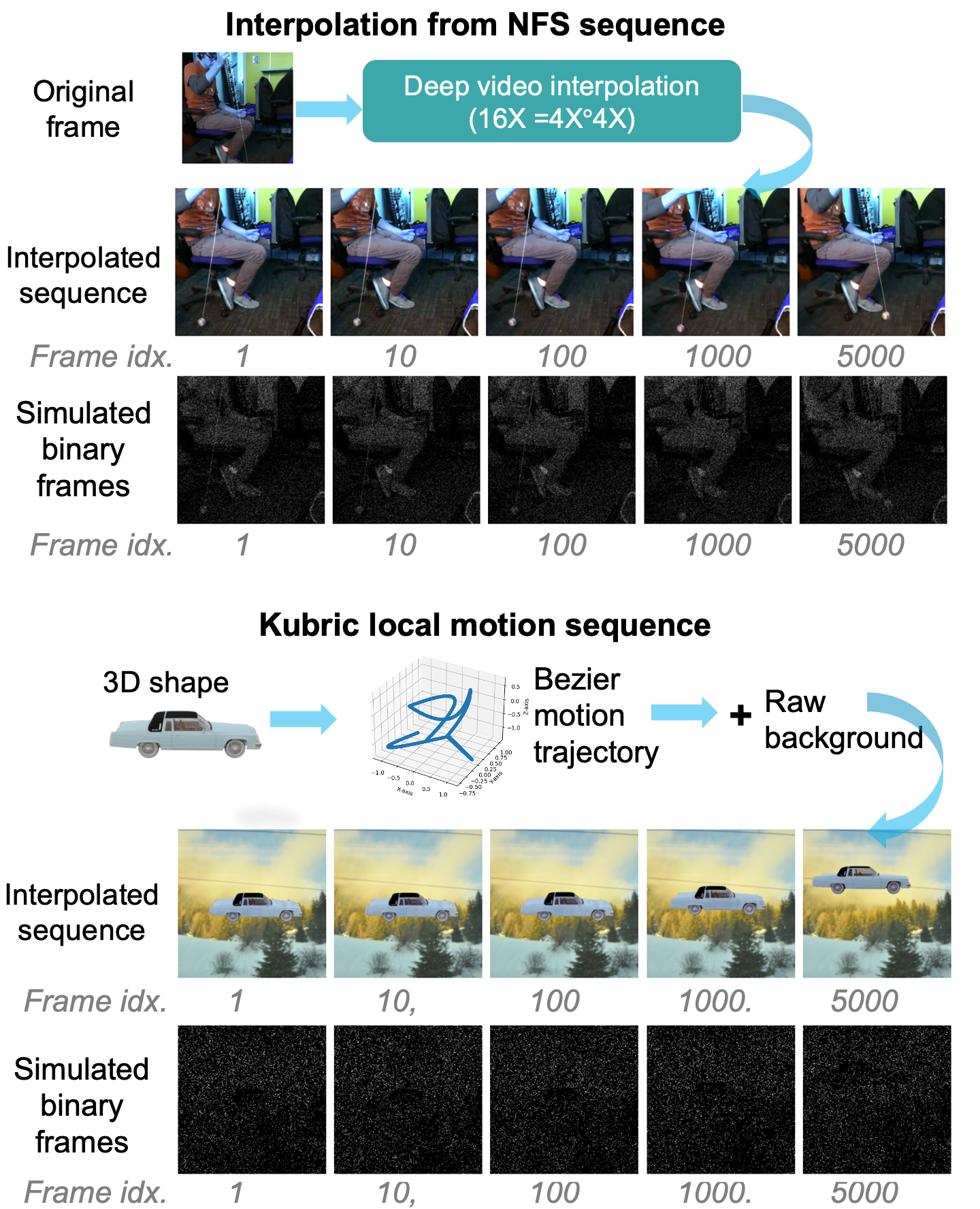}
    \caption{Dataset samples. \textbf{(a)} Real sequence interpolated from the 240~FPS NFS dataset. The interpolated sequence contains fine-grained, realistic motion. The original videos are interpolated 16x, by applying a 4x interpolation twice to avoid artifacts. \textbf{(b)} Kubric samples. We simulate rigid motion of objects (from ShapeNet assets \cite{dang2015raise}) in 3D space with precise control of motion. This allows us to model effects such as perspective change and self-occlusion. We program the foreground with Bezier/linear motion trajectories, add raw static background images \cite{dang2015raise}, and fuse to form a video sequence. Motion sequences produced by (a), (b) can be used to simulate binary frames.}
    \label{fig:dataset_samples}
\end{figure}

\textbf{Clean motion sequences}
First, we produce clean motion sequences without accounting for the exact light level, noise, or sensor characteristics. We expect QIS to operate under a wide range of motion conditions. Hence, it is important to create a rich dataset that covers diverse and realistic objects, motion types, and speeds. 
Our motion sequences are composed of two subsets. For the first subset, inspired by \cite{chandramouli2019bit}, we use RIFE \cite{huang2022real}, a high-performance deep video interpolator, to temporally interpolate the Need for Speed (NFS) dataset \cite{kiani2017need}, which is a high-quality video dataset captured at 240~FPS with negligible motion blur or noise. While the NFS dataset provides realistic textures and motion, it does not cover the entire gamut of local motion speeds and directions. For completeness and generalization, in the second subset we use Kubric \cite{greff2022kubric} to render synthetic local motion sequences, as shown in Fig.~\ref{fig:dataset_samples}~(bottom). The motion sequence set covers a wide range of velocities, objects, occlusions, and motion scenarios. The fine-grained motion is suitable for mimicking underlying intensities at the operation speed of high-speed QIS (0--10000 pixels/s). This diverse dataset is crucial for effectively training the reconstruction network. For more details on the data generation process, please refer to the supplementary document. 

\subsubsection{Quanta sensor model and simulator}
In this section, we discuss how to properly simulate QIS data from motion sequences for training such that the network can adapt to real sensor data.  Given flux values for a sequence of frames, we simulate the binary frame generation process as follows:
\begin{equation}
    B(i,j,t) = N \cdot \text{Bin}(\text{Poisson}(\eta \lambda(i,j,t) + d(i,j))) 
\end{equation}
$B(i, j, t)$ stands for the binary frame at pixel location $(i, j)$ at time $t$. Bin is a binarization operation applied to the photon arrivals; due to pileup: 
\begin{equation}
    \text{Bin}(c) = \begin{cases}
        0, &\quad \text{if } c = 0 \\
        1, &\quad \text{if } c \geq 1
    \end{cases}
\end{equation}
where $c$ indicates the number of photons that arrive during the sampling interval. $\lambda$ is Poisson rate of photon detections. $\eta$ represents the photon detection probability (PDP) and $d$ represents the dark count rate. 

In our work, we use a SwissSPAD device \cite{ulku2018512}, which has relatively uniform PDP and dark count statistics. So here we assume $\eta$ is constant, we collapse $\eta \lambda$ to a single $\lambda$ term. The number of photon arrivals within the duration of a single frame is sampled from a Poisson distribution, with the total rate being $\eta \lambda + d$. We simulate all pixels with a uniform dark count of $d=$7.5 counts per second per pixel. Because the sensor can operate at a rate from 10~kFPS up to 100~kFPS, we scale the per-frame dark count rate to a random amount corresponding to a frame rate in the range of 10--100~kFPS. We calibrate the dark counts from real sensor measurements and pixel locations with significantly higher dark counts are classified as hot pixels. These pixel locations are masked before passing to the neural network. 

In the following paragraphs, we discuss how to find appropriate values for $\lambda$. 

\textbf{Photon flux generation and augmentation}
We now determine the range of flux to simulate when training the NN. The neural network may hallucinate in image regions containing flux values not encountered in the training dataset. To train the network for realistic reconstruction and to avoid hallucination, we need to synthesize data with feasible dynamic ranges. The resolvable dynamic range depends on the number of binary frames \cite{ingle2019high}, as we will also demonstrate below.

\begin{figure}
    \centering
\includegraphics[width=.95\linewidth]{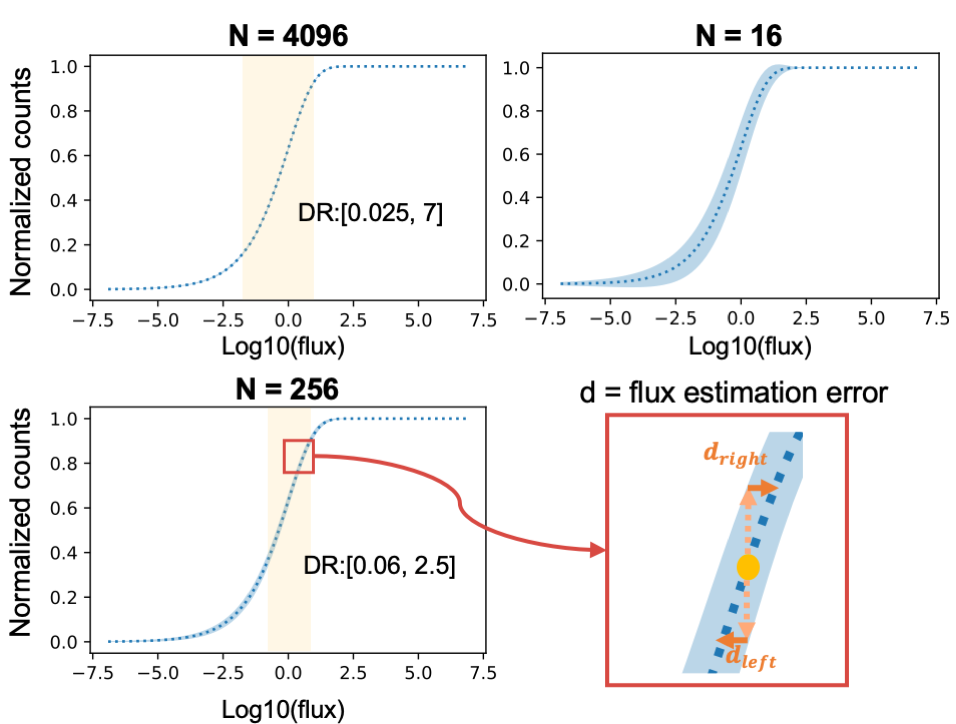}
    \caption{Theoretical dynamic range estimation for bit width $N$. To resolve lower flux and higher flux at the same time (HDR), a higher number of binary measurements is need to be contained in the representation. Using our approach, we plot the forward response curve. We then find the left-most and right-most flux points where the SNR starts to fall below a certain value (showing 20~dB points here). $d_\text{left}$ and $d_\text{right}$ are the left/right flux estimation errors for the given flux. These values dictate the flux ranges to simulate for training data.}
    \label{fig:dynamicrange_characterization}
\end{figure}

What is a realistic limit on dynamic range, given a certain maximum exposure?  We characterize the range by finding flux points with a sufficient amount of SNR or confidence. Let $N$ be the number of frames being captured. As shown in Fig.~\ref{fig:dynamicrange_characterization}, for each $N$ value, we plot the forward response curve, the photon detection probability. Around this line, we also plot the standard deviation of the counts (normalized). These quantities can be computed in closed form, as the number of detected photons follows a binomial distribution. Next, we find flux points in the forward response curve, where the SNR reaches 20~dB and define these points as endpoints of the observable dynamic range. 

The procedure is shown in the bottom right corner of Fig.~\ref{fig:dynamicrange_characterization}. For a flux point $\lambda$, we find the two flux points on the nonlinear curve where the responses match within $p \pm \text{error}$. Let these flux points be $\lambda^-, \lambda^+$. Given a measurement, the flux estimation error is defined as $\epsilon = \text{max}(|\lambda - \lambda^+|, |\lambda - \lambda^-|)$. The SNR is defined as $20 \log_{10}(\lambda / \epsilon)$. For each $N$ value, we find the leftmost and rightmost pair of flux points where the SNR values reach 20~dB. We plot the dynamic range for $N =$ 4096, 256, and 16, where $N$ is the number of binary frames. The results are shown in Fig.~\ref{fig:dynamicrange_characterization}. The larger the value $N$, the more bits are integrated and the wider the dynamic range. For $N =$~4096, the 20~dB flux range is [0.025, 7] per pixel per frame, while for $N =$~256, the range reduces to [0.06, 2.5]. In our experiments, we use $N =$~4096, but we allow the range to extend further to [0.01, 10] due to the denoising ability of neural networks.

\textbf{Piece-wise linear HDR augmentation}
Once we have the above motion sequences, we simulate ultra-fast binary frames with a range of photon fluxes. We perform piece-wise linear HDR augmentation. Given a clean frame in the above sequence, $\lambda$, we augment the fluxes for HDR by thresholding and applying a two-segment piecewise linear transfer function to the clean frame $\lambda$:
\begin{equation}
    \lambda(i,j) = \begin{cases}
        \lambda_{\text{low}} \lambda(i,j) / \max_{i,j}(\lambda),  \quad\quad \text{if} \lambda(i,j) < \lambda_\text{t}\\
        \lambda_{\text{low}} \lambda(i,j) / \max_{i,j}(\lambda) + \\
        \quad (\lambda_{\text{high}}-\lambda_{\text{low}}) (\lambda(i,j) - \lambda_\text{th}) / \max_{i,j}(\lambda)
        , & \text{else.}
    \end{cases}
\end{equation}
$\max(\lambda)$ is the max value in the frame and $\lambda_\text{th}$ is a user-defined threshold. The maximum values in the low- and high-flux regions are $\lambda_{\text{low}}$ and $\lambda_{\text{high}}$. Here we sample $\lambda_{\text{low}}$ $\sim U(0.01, 0.1)$ and $\lambda_{\text{high}}$ $\sim U(0.2, 10)$ during training. We choose a threshold $\lambda_\text{th}$ of around 0.8 (out of 1), such that a small region of pixels is augmented to high flux per sequence. Note that the threshold can also be randomly sampled during training for better generalization. In practice, we found a fixed threshold of 0.8 to be sufficient. The flux-augmented sequences are used as the ground truth references.

\vspace{-1em}
\section{Image/Video Reconstruction Experiments}
We perform image reconstruction experiments on both synthetic and real QIS data. We compare the performance of our approach with a state-of-the-art reconstruction algorithm, QBP~\cite{ma2020quanta}. We also compare with naive integration of the binary frames across various exposure periods.
\vspace{-1em}
\subsection{Training details}
Before the training process, we pre-generate the motion sequences. We randomly split the motion sequences into train, validation, and test subsets. For each sequence, we perform preprocessing, augmentation, and streaming computation of the multi-exposure representation by traversing the video frames in a one-pass manner. We store exposure stacks at 100-bitplane intervals. This process results in a total of 20k training samples. We train the U-Net using the ADAM optimizer with a learning rate of $3 \times 10^{-4}$ and a batch size of 32 for 200 epochs. We randomly crop the samples to size 256x256 and perform random horizontal or vertical flipping (this augmentation is performed online at training time, independent of the data pre-generation).
\vspace{-1em}
\subsection{Synthetic experiments and results}
\begin{figure}[!htbp]
    \centering
    \includegraphics[width=\linewidth]{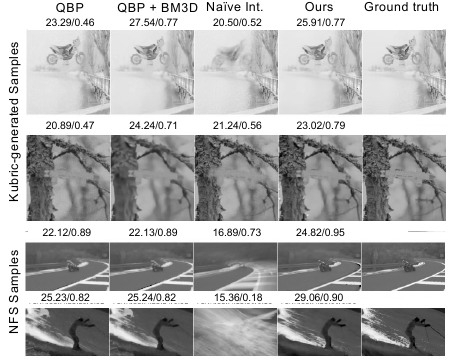}
    \caption{Results on two synthetic test subsets (title for each image corresponds to PSNR/SSIM). \textbf{Top two rows}: synthetic results on the Kubric dataset (results computed with $W_\text{min} =$~16 and $W_\text{max} =$~4096). These scenes are simulated with low light levels, with the \textit{maximum} photon levels of each sample being (from top to bottom) around 0.01~0.1 photons/pixel/frame. \textbf{Bottom two rows}: synthetic results on the NFS dataset. All scenes are simulated to be HDR, with regions as bright as 5--10 photons/pixel/frame, and dark regions with $\ll 1$ photon/pixel/frame. The results show that our method can recover significant details under motion and low light (e.g. motorcycle wheels in the 1st row, car details in the 2nd row, motorcycle in the 3rd row, ski poles in the 4th row) can result in a comparable -- and sometimes better reconstruction than QBP.}
    \label{fig:synthetic_recon}
\end{figure}

We evaluate our methods on the synthetic dataset and compare them with existing approaches. In Table~\ref{tab:results_synthetic}, we report the average PSNR and SSIM across the two data subsets, over a total of 1000 samples. Fig.~\ref{fig:synthetic_recon} visualizes synthetic results on the Kubric and NFS subsets. For QBP, we vary the time-block size and the number of time blocks used for alignment and merging; we found that using a block size of 128 and a block number of 32 gave the best overall results. The QBP results in the figures correspond to using these optimal parameters. For QBP with BM3D denoising, we used $\sigma = 0.05$ for BM3D. BM3D provides a 0--1~dB improvement over QBP alone, depending on how noisy the samples are. We see that our method gives results that are comparable in quality than those produced by QBP, but with lightweight representations and online processing capability.

\begin{table*}[!hbtp]
\centering
\caption{Quantitative Results on the Synthetic Dataset}
\label{tab:results_synthetic}
\begin{tabular}{|l|c|c|c|c|c|c|c|c|}
    \hline
    \multirow{2}{*}{Motion subset} & \multicolumn{2}{c|}{QBP} & \multicolumn{2}{c|}{QPB+BM3D ($\sigma=0.05$)} & \multicolumn{2}{c|}{Naive integration} & \multicolumn{2}{c|}{Ours} \\ \cline{2-9}
    & PSNR & SSIM & PSNR & SSIM & PSNR & SSIM & PSNR & SSIM \\ \hline
    Kubric local motion & $23.74 \pm 1.78$ & $0.58 \pm 0.14$ & $25.22 \pm 1.21$ & $0.68 \pm 0.12$ & $22.10 \pm 2.11$ & $0.64 \pm 0.12$ & \textbf{25.54 $\pm$ 2.21} & \textbf{0.80 $\pm$ 0.05} \\ \hline
    NFS interpolated & $31.08 \pm 6.56$ & $0.93 \pm 0.08$ & \textbf{31.10 $\pm$ 6.57} & $0.93 \pm 0.08$ & $23.10 \pm 7.73$ & $0.75 \pm 0.24$ & $28.60 \pm 7.71$ & \textbf{0.95 $\pm$ 0.04} \\ \hline
\end{tabular}
\end{table*}

\vspace{-1em}
\subsection{Real data}
\begin{figure*}[!htbp]
    \centering
    \includegraphics[width=\textwidth]{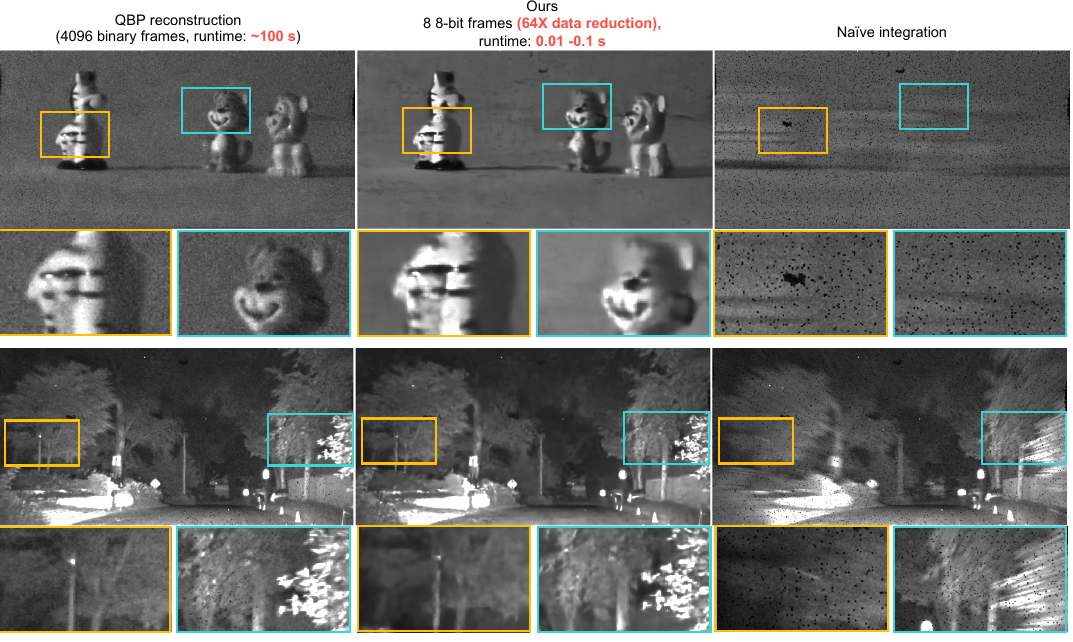}
    \caption{Reconstruction on real SPAD-QIS scenes containing global motion. \textbf{Top:} A scene containing toys on a table, shot under ambient indoor lighting. \textbf{Bottom:} A night driving scene, shot from a moving car. Both QBP and our method results in sharp images, even under dark lighting conditions and fast motion. Compared to QBP, our method gives significantly reduced bandwidth and computation time. In both scenes, our method reconstructs details well (patterns on the toys and leaves). The naive integration also uses 4096 binary frames ($\approx$ 0.3s), and results in large motion blur due to rapid motion.}
    \label{fig:real_recon_global}
\end{figure*}

\begin{figure*}[!htbp]
    \centering
    \includegraphics[width=\textwidth]{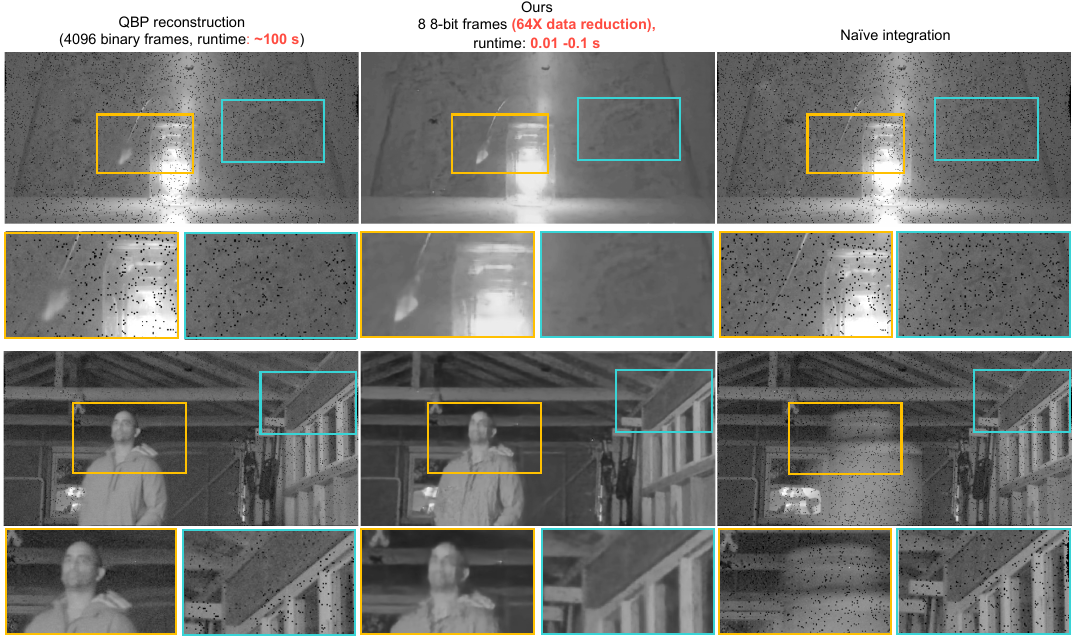}
    \caption{Reconstruction on real SPAD-QIS scenes containing local motion. \textbf{Top:} This scene consists of a very dark background with a bright lamp in front, which covers a large dynamic range. A small pendulum swings back and forth in front of the light. We reconstruct the moment when the pendulum almost reaches its maximum speed. Our approach resolves the pendulum as well as the background. \textbf{Bottom:} This scene contains a person walking. Our approach results in a good face reconstruction with visible facial landmarks. In both cases, the moving and static portions are both reconstructed well.}
    \label{fig:real_recon_local}
\end{figure*}

Additionally, we apply reconstruction to a real QIS dataset introduced in \cite{ma2020quanta} and \cite{ma2023burst}. Similar to the simulation process, we mask the hot pixels based on calibrated hot-pixel masks. The real data comes directly in binary streams, hence we directly extract the streaming representation on the binary data. We apply the trained U-Net to the streaming representations and show the resulting reconstruction. We visually compare the results with QBP and naive integration. The evaluation is purely qualitative, given the lack of ground-truth intensity data. Fig.~\ref{fig:real_recon_global} and Fig.~\ref{fig:real_recon_local} show reconstructions scenes with global and local motion, respectively. The scenes are recorded in challenging settings with significant motion and very low light ($\ll$ 1 photon/pixel/frame). Overall, our results are comparable to QBP, and despite large motion or low light, our method produces sharp, high-quality reconstructions in real-time.

\vspace{-1em}
\subsection{Runtime comparison}
Quanta burst photography takes 1.5--2 minutes to run on an Intel Xeon E3-1220 v2 CPU for each reconstruction (4096 frames). Here to be fair, we used a high-end GPU resulting in the QBP speed to be 10X reported by the original paper \cite{ma2020quanta}. However, even with the speedup it's far from real-time. Furthermore, with QBP, the reconstruction time scales with the number of binary frames used per reconstruction. With further optimization, the speed may be improved, but it may still be challenging to achieve real-time performance, especially for higher-resolution sensors. In contrast, our method runs at about 12~ms per reconstruction on an A100 GPU, and can still easily achieve tens of FPS on lower-end, desktop-level GPUs. Because the input always consists of 8 channels, runtime is independent of the exposure period.

\vspace{-1.4em}
\subsection{Characterizations using synthetic data}
The goal of this section is to study how the performance of our algorithm varies as the motion level and light level change. To have tight control over speed and light levels, we simulate two scenes, each programmed with a distinct type of motion:
(1) a scene with high-frequency textures with global motion, and (2) a local translation of a cow with distinct patterns. We vary the motion speed and light level for the two scenes. We assume a frame rate of 20~kFPS.

\begin{figure*}[!htbp]
    \centering
    \includegraphics[width=\textwidth]{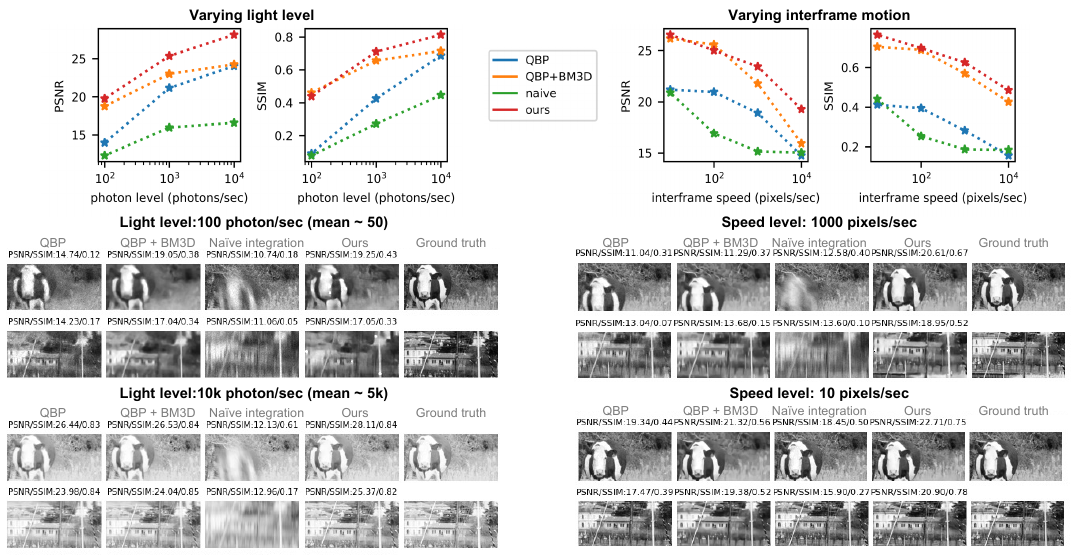}
    \caption{Performance under various speeds. \textbf{Left side:} The effect of varying light levels (values give the maximum photon level per sequence). \textbf{Right side:} The effect of varying inter-frame speed, with the light level being fixed to around 200 photons/pixel/second on average. \textbf{Top-row plots:} For the PSNR/SSIM plots (top), we render 5 continuous motion sequences for each light and motion configuration. We obtain a total of 250 samples per light/motion point by taking 50 samples across each sequence and performing reconstructions. Our method results in comparable or better performance than QBP across different lighting and motion configurations.}
    \label{fig:char_speed_summary}
\end{figure*}

\textbf{Effects of light level:}
To characterize the effect of light level, we fix the speed to 1000 pixels/second. We vary the maximum light level between [100, 1k, 10k] photons per second. Both 50--100 and 500--1000 photons per second would be considered low light. We show our results in Fig.~\ref{fig:char_speed_summary}; our approach results in the overall best PSNR and SSIM.

As the light level becomes smaller, all the methods have lower performance. At very low light, our method and QBP with BM3D denoising show the highest reconstruction qualities. Overall, the two approaches are comparable in terms of quality; at 100 photons per second, both methods can resolve some scene texture, but their performance is heavily limited by the low photon numbers. Both give clean results at 1000 photons per second, with our method a bit better at preserving high-frequency details.\\
\textbf{Effects of motion speed:}
For the simulated scenes, we translate a raw image frame around so that it stays in the field of view. We adjust the translation speed of the scene or object within a range of [100, 1000, 10000] pixels per second, corresponding to an inter-frame motion of [0.005, 0.05, 0.5] pixel/frame. We use a fixed photon rate across the various motion speeds. The maximum rate is about 500 photons/second/pixel, with the average being about 200 photons/second/pixel. Our method results in similar performance to QBP, with the performances worsening as speed increases. Beyond 1000 pixel/s, QBP cannot properly align to the latest intensity frame (e.g., see the right column of Fig.~\ref{fig:char_speed_summary}). This might be due to failed optical flow near the end of the sequence.   

\vspace{-1em}
\section{Vision Applications}

We show that the proposed system can be applied to real-time computer vision (tens of FPS), by applying off-the-shelf vision inference models on reconstructed images. Later, we discuss the possibility of applying end-to-end training in various tasks.
\vspace{-1em}
\subsection{Object detection (20~FPS)}
\begin{figure*}[!hbpt]
    \centering
    \includegraphics[width=\linewidth]{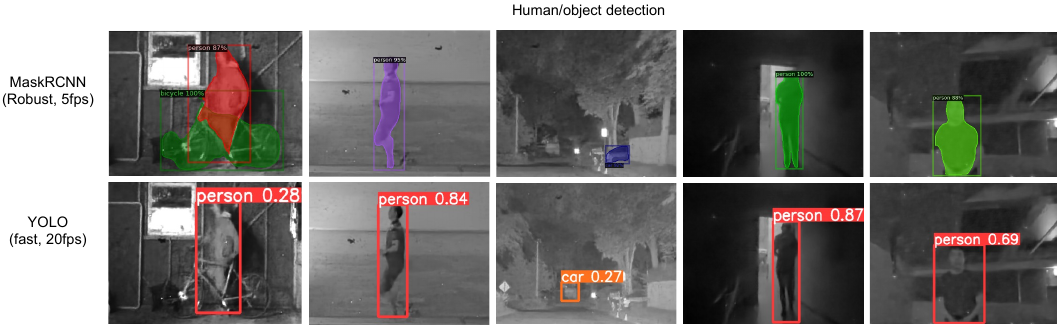}
    \caption{Detection results on real data, using MaskRCNN and YOLO. All of the scenes shown here were captured in low light -- from dark rooms to streets at night, with subjects/cameras moving at a significant speed. The reconstructions from our method are of sufficient quality that we can directly apply off-the-shelf detection algorithms to the reconstructions. We achieve good detection results, even under the challenging imaging conditions shown here.}
    \label{fig:detection}
\end{figure*}

Fig.~\ref{fig:detection} shows results for object detection. Here we apply two state-of-the-art networks: MaskRCNN \cite{he2017mask} and YOLOv5\cite{redmon2016you,yolov5}. MaskRCNN tends to be more robust (as seen in the first row of Fig.~\ref{fig:detection}) but slower, running at 5~FPS. We used YOLO v5, which has a runtime of 20~FPS, but gives less stable predictions. The image reconstructions can directly pair with these networks, and we can optimize our network configuration depending on quality vs. speed requirements. 

\vspace{-1em}
\subsection{Human-interaction tasks: face and pose detection}
\begin{figure}[h]
    \centering
    \includegraphics[width=.8\linewidth]{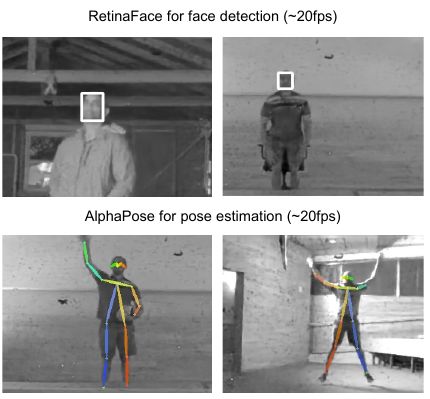}
    \caption{Human-interaction tasks including face detection and pose estimation on real QIS captures.  All scenes shown here were captured under low light. Thanks to the compute- and bandwidth-efficient nature of our method, these tasks can be performed in real-time, even when the subjects have fast motion and are subject to low illumination.}
    \label{fig:human}
\end{figure}

For applications like AR/VR and human-computer interaction, face detection and pose estimation are important tasks with a requirement to run in real-time. In Fig.~\ref{fig:human}, we show a few examples of applying face detection (RetinaFace~\cite{deng2020retinaface}) and pose estimation (AlphaPose~\cite{fang2022alphapose}). Both tasks achieve good quality and run at 20~FPS. 

\begin{figure*}[h]
    \centering    
    \includegraphics[width=\linewidth]{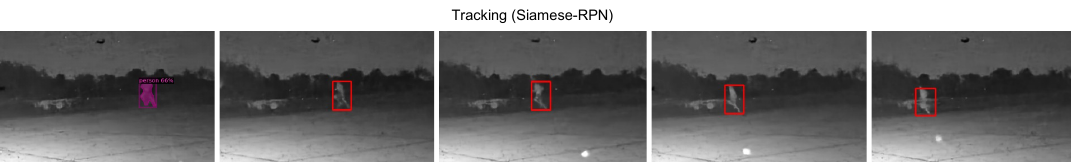}
    \caption{Tracking results (Siamise-RPN) on a scene containing a running person, captured outdoors at night. We see reliable tracking results, despite the challenging nature of this scene. These results further highlight our method's ability to recover high-quality images and to support a range of downstream vision tasks.}
    \label{fig:track}
\end{figure*}
\vspace{-1em}
\subsection{Tracking}
Here we demonstrate tracking using Siamese-RPN \cite{li2018high} on reconstructions from our method. Results are shown in Fig.~\ref{fig:track}. The initial template is defined as the human detection (pink) in the first frame. For the later reconstructions, we track the position of the human (red box).

\section{Discussion}
\vspace{-1em}
\subsection{Limitations}
While our approach achieves good-quality reconstructions, methods such as QBP that are optimized for maximizing image quality may achieve higher fidelity for fine details. Our work is also not targeted for extremely high flux. However, our framework can be extended to incorporate high flux values during data synthesis and training. Furthermore, since our reconstruction is based on supervised neural networks, it can sometimes contain artifacts, and the performance is influenced by the training data distribution. In application cases such as face recognition or bar code scanning, where a high degree of fidelity on high-frequency information is needed, QBP is likely to provide more reliability in fine details. However, our approach is more suitable where data bandwidth and real-time performance are crucial. 
\vspace{-1em}
\subsection{End-to-end vision}
Going beyond applying off-the-shelf networks to reconstructed frames, one could also study end-to-end training of the network using the multi-exposure representation directly. Future work can apply similar data creation procedures to those discussed in this paper, using existing datasets and data produced with graphics toolboxes. One option for tracking is the NFS dataset, which not only provides videos but also bounding boxes at each frame. To produce paired streamed representation with the label, one can first interpolate the video and then generate streamed representations on the interpolated video at a step size that is a multiple of the interpolation factor, such that each frame of the resulting representation corresponds to a labeled frame. Other data generation options are graphics frameworks such as Kubric and PointOdyssey~\cite{zheng2023pointodyssey}, which can produce close-to-realistic, high-quality simulated data. Their outputs could include intensity sequences along with various labels such as depth or optical flow. Segmentation and tracking labels can also be easily generated as the object type and trajectory are known ahead of time. PointOdyssey also already provides a large labeled video dataset, containing not only objects and assets but also human avatars.
\vspace{-1em}
\subsection{Other streaming representations and inference architectures}
There are also other possibilities in streaming representations, e.g., higher-order statistics, projections, or codings \cite{sundar2023sodacam}. One can also consider integrating with differential approaches \cite{white2022differential}\cite{zhang2022first} to achieve higher bit-efficiency. Multi-spatial resolutions may also be exploited to reduce the data bandwidth further. Our work currently considers grayscale flux, but the streaming framework also applies to the acquisition of color images as well (i.e., by attaching a Bayer filter). 
\vspace{-1em}
\section{Acknowledgements}
T.Z., V.B., A.V. were supported in part by the NSF Expeditions IIS-1730574 and NSF Computational Thermal Award IIS-2107313. M.G. and M.D. were supported in part by the NSF CAREER Award 1943149, and in part by the NSF award under Grant CNS-2107060. We would also like to thank Sizhuo Ma for providing the QBP code and useful discussions, and Varun Sundar for helping with data preparations.

\bibliographystyle{IEEEtran}
\bibliography{references}

\begin{IEEEbiography}[{\includegraphics[width=1in,height=1.25in,clip,keepaspectratio]{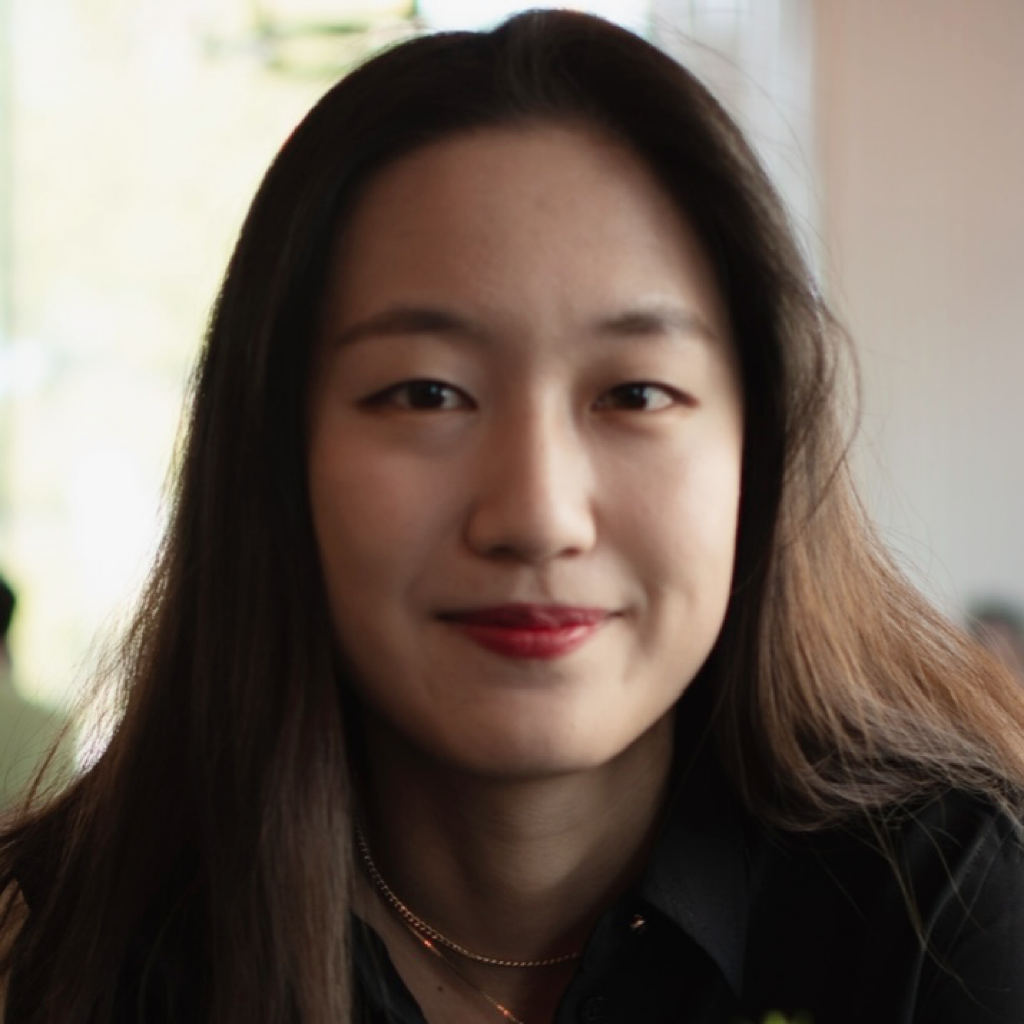}}]{Tianyi Zhang} is currently a Ph.D student at Rice University, where she also received her M.S. degree in 2022, B.S. degree (Hons.) in 2019 in electrical engineering. Her research interests include exotic image sensors, SPADs, quanta image sensors, computational imaging and computer vision.
\end{IEEEbiography}

\begin{IEEEbiography}[{\includegraphics[width=1in,height=1.25in,clip,keepaspectratio]{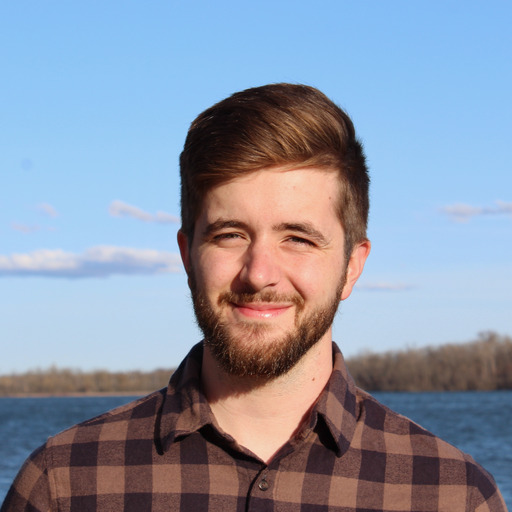}}]{Matthew Dutson}
is currently a Ph.D. student in the Computer Sciences department at the University of Wisconsin--Madison, where he received his M.S. degree in 2020. Previously, he earned a B.S. degree in physics from the University of Utah (2018). His research interests broadly include computer vision and computational imaging, with a particular focus on resource-efficient sensing and inference.
\end{IEEEbiography}

\begin{IEEEbiography}
[{\includegraphics[width=1in,height=1.25in,clip,keepaspectratio]{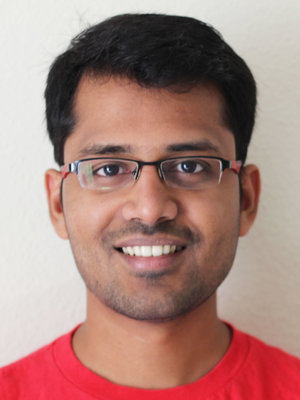}}]{Vivek Boominathan} received the BTech degree in electrical engineering from the Indian Institute of Technology Hyderabad, Hyderabad, India, in 2012, and the MS and Ph.D. degrees from the Department of Electrical and Computer Engineering, Rice University, Houston, Texas, in 2016 and 2019, respectively. He is currently a postdoctoral associate with Rice University, Houston, Texas. His research interests include the areas of computer vision, signal processing, wave optics, and computational imaging.
\end{IEEEbiography}

\begin{IEEEbiography}[{\includegraphics[width=1in,height=1.25in,clip,keepaspectratio]{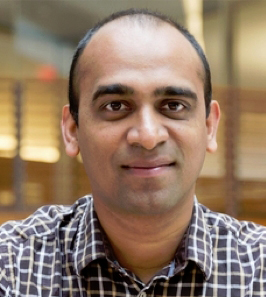}}]{Mohit Gupta}
is an Associate Professor of Computer Sciences at the University of Wisconsin-Madison. He received Ph.D. from the Robotics Institute, Carnegie Mellon University, and was a postdoctoral research scientist at Columbia University. He directs the WISION Lab with research interests broadly in computer vision and computational imaging. He has received the Marr Prize honorable mention at IEEE ICCV, a best paper honorable mention at IEEE ICCP, a Sony Faculty Innovation Award and an NSF CAREER award. 
\end{IEEEbiography}

\begin{IEEEbiography}[{\includegraphics[width=1in,height=1.25in,clip,keepaspectratio]{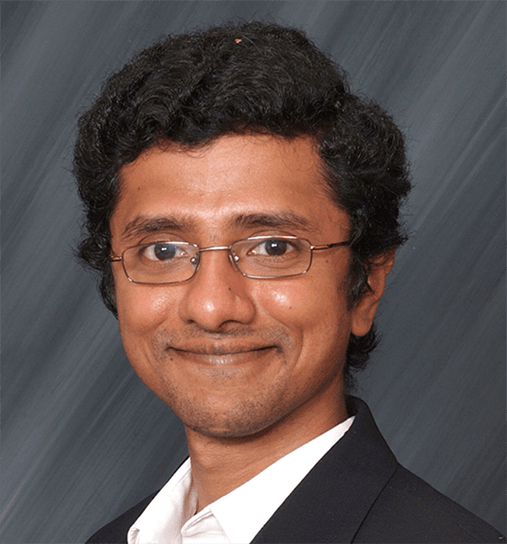}}]{Ashok Veeraraghavan}
received the B.S. degree in electrical engineering from the Indian Institute of Technology, in 2002, and the M.S. and Ph.D. degrees from the Department of Electrical and Computer Engineering, University of Maryland, in 2004 and 2008, respectively. He is currently a Professor of electrical and computer engineering at Rice University. Before joining Rice University, he spent three years as a Research Scientist with Mitsubishi Electric Research Labs, Cambridge, MA, USA. His research interests include computational imaging, computer vision, machine learning, and robotics. His thesis received the Doctoral Dissertation Award from the Department of Electrical and Computer Engineering at the University of Maryland. He received the NSF CAREER Award in 2017.
\end{IEEEbiography}


\end{document}